\documentclass[12pt,preprint]{aastex}

\shorttitle{Luminosities and Space Densities of Short GRBs}
\shortauthors{Schmidt}

\begin{document}

\title{Luminosities and Space Densities of Short Gamma-Ray Bursts}

\author{Maarten Schmidt}
\affil{California Institute of Technology, Pasadena, CA 91125}
\email{mxs@deimos.caltech.edu}

\begin{abstract}

Using the Euclidean value of $<V/V_{\rm max}>$ as a cosmological
distance indicator, we derive the isotropic-equivalent characteristic
peak luminosity of gamma-ray bursts both longer and shorter than
2 s. The short bursts have essentially the same 
characteristic peak luminosity of
$0.6 \times 10^{51}$ erg (0.064s)$^{-1}$ as do the long bursts. This 
may apply also to bursts with durations less than 0.25 s.
The local space density of short bursts is around three times lower
than that of long bursts.

\end{abstract}


\section{Introduction}

Since 1997, our understanding of gamma-ray bursts (GRB) has increased
enormously through the optical identification of afterglows, determination
of redshifts from optical spectra and generally from afterglow studies 
covering a large range of energies, from radio to X-rays. These studies
suggest that GRBs are associated with massive stars and hence that the
burst rate may be linked to the rate of star formation in galaxies.
The number of GRBs so observed is still rather limited, and those 
successfully studied are all long bursts. We use the classification 
of GRBs in long bursts and short bursts based on the
distribution of durations $T_{90}$ in the BATSE data \citep{kou93},
which show a minimum around 2 s.  

No afterglows have been observed so far for short 
bursts \citep{gan00}. As a consequence, there is no
direct knowledge of their redshifts, luminosities and space densities.
There has been a suspicion that short GRBs would be at smaller redshifts
and be of lower luminosities than long bursts, perhaps due to a remark
in the abstract of a paper by \citet{tav98} that short bursts show
little if any deviation from a Euclidean distribution but this
is not supported by data discussed in the paper.

In this Letter, we briefly discuss
efforts to derive the luminosity function of long bursts and then
present a derivation of the characteristic peak luminosity $L^*$ 
and the local space density for both long and short bursts.
Even though the total gamma-ray energy may be physically of
greater interest \citep{fra01}, we concentrate on
peak luminosities since the detection of GRBs is based on count 
rates, not on the time-integrated flux or fluence of the burst.

Before any redshifts of GRBs were known, \citet{mao94} used the 
Euclidean value of $<V/V_{\rm max}>$ as a relative distance
indicator to show that long and short bursts had the same peak
luminosity to within a factor of 2, assuming they were
standard candles. The redshifts that are now 
available for long bursts have led to the development of 
several luminosity indicators, such as the spectral lag derived from
cross-correlation of two spectral channels \citep{nor00} and
the variability in the time profile \citep{fen00}. These
luminosity indicators can in principle be used to derive the
luminosity function of GRBs, including its evolution with redshift.
\citet{nor01} have shown that the spectral lags for short bursts
are much smaller than those of long bursts and conclude that the
lag magnitude is discontinuous accross the 2 s valley between long
and short bursts.

We have used the Euclidean value of $<V/V_{\rm max}>$ as a 
cosmological distance indicator to first derive characteristic
luminosities \citep{sch99b} and then the luminosity function of GRBs
\citep{sch01}. This method, which makes use of a large sample of GRBs,
circumvents the current weakness of methods based on the small
number of redshifts observed so far. The price we pay for using
$<V/V_{\rm max}>$ as a distance indicator is that an assumption 
has to be made about the evolution of the luminosity function with 
redshift. We assume that this evolution tracks the star formation
rate (SFR) on the expectation that GRBs are associated with massive
stars. We employ the parametrizations of \citet{por01}, in particular
their SFR model SF2 in which the co-moving space density
rises by an order of magnitude near redshift
$z=1$ and then remains roughly constant for $z>2$ \citep{ste99}.

Our application of $<V/V_{\rm max}>$ as a distance indicator 
used the BD2 sample (see Sec. 2) which is based on BATSE DISCLA data
on a time resolution of 1024 ms. The resulting characteristic
luminosities \citep{sch99b} and luminosity functions \citep{sch01} 
therefore applied to GRBs with a duration exceeding 1 or 2 s. 
In this Letter we employ the same method to derive characteristic
luminosities $L^*$ using the BATSE catalog.
By using the GRBs detected with a time resolution of 64 ms,
we can derive the $L^*$ of short bursts.
We will actually employ time resolutions of both 64 and 1024 ms, and
obtain $L^*$ for both short and long bursts, so that we can directly
compare them.

In Sec. 2, we review the application of $<V/V_{\rm max}>$ in
deriving the luminosity function of long GRBs from the BD2 sample
without using redshifts and recall the main results obtained.
Following a discussion of data from the BATSE catalog in Sec. 3, 
we derive characteristic luminosities for both long and short GRBs
in Sec. 4. The results are discussed in 
Sec. 5. Throughout this Letter, we use a flat cosmological model
with $H_0 = 65~$ km s$^{-1}$ Mpc$^{-1}$, $\Omega_M = 0.3$, and
$\Omega_{\Lambda} = 0.7$.

\section{Luminosity Function Of Long Bursts Derived From The BD2 Sample}

The BD2 sample of GRBs is based on BATSE DISCLA data consisting of
the continuous data stream from the eight BATSE LAD detectors in
four energy channels on a timescale of 1024 ms \citep{fis89}. The
sample was derived using a software trigger algorithm requiring an
excess of at least 5$\sigma$ over background in at least two detectors
in the energy range $50-300$ keV. The initial version was described in
\citet{sch99a} and a revision in Sec. 2 of \citet{sch99b}. The BD2 sample
covers a period of 5.9 yr from TJD $8365-10528$. It contains 1391 GRBs
of which 1013 are listed in the BATSE catalog. 
The value of $<V/V_{\rm max}> = 0.336\pm0.008$. The sample of 1391 GRBs
effectively represents 2.003 yr of full sky coverage, corresponding to
an annual rate of 694 GRBs.

The derivation of the luminosity function of GRBs \citep{sch01} was 
based on the correlation of the Euclidean value of $<V/V_{\rm max}>$ 
with spectral hardness. Given that the Euclidean value
of $<V/V_{\rm max}>$ for a well defined sample of cosmological
objects is a distance indicator, we interpreted the correlation of 
$<V/V_{\rm max}>$ with spectral hardness as a luminosity-hardness
correlation. The luminosity function 
was derived as the sum of the luminosity functions of four 
spectral hardness classes. It can be characterized approximately
as consisting of two power laws of slopes $-0.6$ and $-2$, respectively,
with an isotropic-equivalent
break peak luminosity of $\log L^* \sim 51.5$. The luminosity
function ranges approximately from $\log L^*-1.5$ to $\log L^*+1.0$.

In the derivation of the characteristic luminosity $L^*$ of GRBs 
from BATSE data
in Sec. 4, we will assume that the shape and extent of the luminosity
function is that of the broken power law just described and
derive the value of $\log L^*$ from $<V/V_{\rm max}>$. This is
essentially the method used in \citet{sch99a}, where we varied the
assumed extent and shape of the luminosity function to study the 
effect on the derived value of $\log L^*$.

\section{Using Data From The BATSE Catalog}

\subsection{Evaluating $V/V_{\rm max}$}

We will be using $<V/V_{\rm max}>$ values derived from the BATSE
4B catalog\footnote{See 
http://gammaray.msfc.nasa.gov/batse/grb/catalog/4b catalog 
maintained by W. S. Paciesas et al.}
for bursts both longer and shorter than 2 s. Before we
apply $<V/V_{\rm max}>$ as a cosmological distance indicator, we
consider how the individual $V/V_{\rm max}$ values in the
BATSE catalog are derived, and also how they are affected by the 
imposition of duration limits.

The BATSE catalog lists for individual bursts the count rate in
the second brightest illuminated detector, $C_{max}$, as well as
the minimum detectable rate $C_{min}$. The value of $V/V_{\rm max}$
is then simply derived as $(C_{max}/C_{min})^{-3/2}$. In contrast,
the values of $V/V_{\rm max}$ in the BD2 sample have been derived
through simulations, in which the burst is moved out in Euclidean
space in small steps with a corresponding reduction in its amplitude.
At each step, the full detection algorithm is re-employed to set the
background and detect the burst. Once the burst is not detected any more,
the value of $V/V_{\rm max}$ is simply derived from the reduction factor.
During this process of removal, the burst may get detected later and 
later depending on the time profile. The background time window, 
which precedes the detection by a fixed time interval, may start to 
include some burst signal. The final detection is usually made on
the peak of the burst but in some cases where the burst signal
preceding the peak is high and enters the background, the final
detection may be off the peak. Given that in these 
cases upon removal the burst drops out earlier than
expected from $C_{max}/C_{min}$, the actual value of $V/V_{\rm max}$ 
will be larger. In practice, the effect depends much on the time
profile. The net effect for a sample of bursts is that
$(C_{max}/C_{min})^{-3/2}$ is an underestimate of $<V/V_{\rm max}>$.

Next we consider the effect of imposing duration limits, such as
$T_{90} <$ 2 s. In the simulations carried out on the BD2 sample
described above, we found that the duration of the burst decreased
as it was moved out until at the limit of detection it was 1 or 2 s.
A qualitatively
similar effect has been described as a fluence duration bias for GRBs 
in the BATSE catalog \citep{hak00}. Suppose our sample is set by a 
restriction involving a limiting duration $T_{lim}$. In deriving 
$V/V_{\rm max}$, we should strictly apply two simultaneous limits, 
namely $C_{min}$ and $T_{lim}$, as was done in the first $V/V_{\rm max}$
application \citep{sch68}. However, evaluating the effect of $T_{lim}$ 
on $<V/V_{\rm max}>$ would require simulations of the derivation of 
$T_{90}$ for BATSE GRBs which are not available.
Therefore, we limit ourselves to finding the sign of the systematic 
error in $<V/V_{\rm max}>$ if we ignore $T_{lim}$.

Consider the case of a GRB with a duration $T_{90} > T_1$.
As we move the burst out, $T_{90}$ will decrease and may become
smaller than $T_1$ before it becomes undetectable. Therefore
the reduction factor is smaller and $V/V_{\rm max}$ is larger.
Ignoring the lower limit $T_1$ leads to an underestimate of 
$V/V_{\rm max}$. In the case of an upper limit $T_2$, a GRB with 
$T_{90} > T_2$ which does not belong to the sample, may become part
of it when its $T_{90}$ becomes shorter than $T_2$ upon removal.
In this case, ignoring the upper limit $T_2$ produces
a $V/V_{\rm max}$ that is an overestimate. It should be emphasized that
these considerations are purely qualitative; the actual effects
depend on such factors as the time profile of the burst, the way
the background is set, etc.

\subsection{Effective Coverage}

In order to derive the rate of GRBs per unit volume, we need to have 
an estimate of the effective full sky coverage of the GRB sample used. 
For the BD2 we have evaluated the efficiency at $33.8\%$ leading
to an effective full sky coverage of 2.003 yr \citep{sch99a}. 
The total sample of 1391 GRBs then corresponds to a rate of 
694 GRB yr$^{-1}$.

The 4B catalog gives an annual rate of 666 GRB yr$^{-1}$,
presumably based on the most sensitive detections, which are at 
the 1024 ms timescale. Considering that the S/N limit of the 4B  
catalog is 5.5$\sigma$ and that of the BD2 sample 5$\sigma$,
these rates are quite consistent. We only consider GRBs
detected while the BATSE on-board trigger was set for 5.5$\sigma$ 
over the energy range $50-300$ keV. Among those, the 4B catalog 
contains 612 GRBs for which $C_{max}/C_{min} \ge 1$ at the 
1024 ms timescale (those $ < 1$ were detected at time scales of
64 ms or 256 ms). This corresponds to an effective full sky 
coverage for the purpose of this work of 0.92 yr. We assume 
that this value also applies to the 64 ms timescale.

\section{Characteristic Luminosities For Long and Short Bursts}

The method used to derive the isotropic-equivalent 
characteristic peak luminosity $L^*$
for a given value of $<V/V_{\rm max}>$ is essentially the same
as that used before \citep{sch99b}. Based on our discussion of
long bursts in Sec. 2, the local luminosity function of peak 
GRB luminosities $L$, defined as the co-moving space density of
GRBs in the interval $\log L$ to $\log L + d\log L$, is
\begin{mathletters}
\begin{eqnarray}
\Phi_o(L) & = & 0, \qquad\hbox{for}\qquad \log L < \log L^* - 1.5, \\
\Phi_o(L) & = & c_o (L/L^*)^{-0.6}, \qquad\hbox{for}\qquad
 \log L^* - 1.5 < \log L < \log L^*, \\
\Phi_o(L) & = & c_o (L/L^*)^{-2.0}, \qquad\hbox{for}\qquad
 \log L^* < \log L < \log L^* + 1.0, \\
\Phi_o(L) & = & 0, \qquad\hbox{for}\qquad \log L > \log L^* + 1.0.
\end{eqnarray}
\end{mathletters}
We assume that the GRB rate as a function of redshift follows 
the SFR model SF2 (see Sec. 1). The median value of the
spectral photon index for the (long) bursts in the BD2 sample is
$-1.6$ \citep{sch01}. For the short bursts we adopt an index of $-1.1$
to reflect their larger average hardness ratio \citep{kou93}.

The median 4B limiting photon flux for 1024 ms detection in the 
$50-300$ keV range, including the effect of atmospheric scattering, is 
0.25 ph cm$^{-2}$ s$^{-1}$ \citep{mee98}. We adopt the same value
for the BD2 sample for which atmospheric scattering has not yet been
evaluated. For GRBs detected on the 64 ms timescale, the derivation
of $L^*$ from $<V/V_{\rm max}>$ is carried out on this timescale,
so the resulting luminosity $L^*$ is produced in ergs (0.064s)$^{-1}$. 
In order to convert to this system from a timescale of 1024 ms, 
we compared for GRBs with $T_{90} > 2$ s their peak fluxes
at timescales 1024 ms and 64 ms given in the BATSE catalog.
The average 1024/64 flux ratio is 0.68, reflecting the variability 
of long GRBs at their peak at subsecond timescales. Accordingly, we 
convert peak luminosities per 1024 ms into ones per 64 ms by dividing
by $ 0.68 \times 16 = 10.9$. 

In order to provide a comparison with past work, we first derive
$L^*$ for long bursts. The top line of Table~\ref{tbl-1} shows the
results for the BD2 sample of GRBs. The value of $L^*$ is 70\%
larger than that found from the more sophisticated 
derivation in \citet{sch01}, which involved splitting the sample in groups 
of different spectral hardness. The resulting luminosity function was
not exactly a broken power law, which causes the above difference.

The remaining entries in Table~\ref{tbl-1} are all based on data
from the 4B catalog. The first three can be directly compared
with the BD2 results since they all concern long bursts. The
$\log L^*$ values show a range of 0.3, with a systematic offset 
of around $+0.2$ from the BD2 value. The local densities
are systematically lower than that from the BD2, partly due
to the higher luminosity.

The next row of Table~\ref{tbl-1} gives the results for short bursts
with $T_{90} < 2$ s.
The characteristic peak luminosity $L^*$ is in the middle of the
range given for long BATSE bursts. The local space density of
short bursts is $\sim 3$ times smaller than that of long bursts.  
In order to investigate whether there might be a trend among the short 
bursts, we show further results for $T_{90} < 0.5$ s and $< 0.25$ s.
The formal mean errors in $\log L^*$ for the short bursts are
$\pm 0.20$, $\pm 0.25$, and $\pm 0.35$, respectively. We see no
trend of $\log L^*$ with duration among the short bursts.

\section{Discussion}

The systematic effect of using $C_{max}/C_{min}$ in the derivation
of $<V/V_{\rm max}>$ discussed in Sec. 3 applies to all BATSE
entries in Table~\ref{tbl-1}. Therefore all values of $<V/V_{\rm max}>$
are systematically too small, and the resulting values of $\log L^*$
too large. The actual effect is hard to estimate because it depends
on the burst profiles and the methodology of setting the background.
For the BD2 sample we find that using $C_{max}/C_{min}$ results in
an underestimate of $<V/V_{\rm max}>$ of 0.009 leading to an
overestimate of $\log L^*$ by 0.08. Considering this offset the
agreement between the first two rows of Table~\ref{tbl-1} is
satisfactory considering that the formal errors (derived from the 
mean errors of $<V/V_{\rm max}>$) in $\log L^*$ are $\pm 0.07$ and 
$\pm 0.10$, respectively. For the BATSE bursts with $T_{90} > 2$ s 
in the next two rows of Table~\ref{tbl-1}, the systematic effect 
caused by the duration limit may result in an overestimate of 
$\log L^*$, but no obvious effect is evident. 

The short bursts with $T_{90} < 2$s have a value of $\log L^*$
that is entirely consistent with that of the long bursts. The
systematic effect of the upper limit for the duration will be
an underestimate of $\log L^*$. This effect may well be
small for short bursts, but only simulations can tell.

The peak luminosities and local space densities given in 
Table~\ref{tbl-1} are isotropic-equivalent values and are all based 
on SFR model SF2 (Sec. 1). If, instead, we use SFR models SF1 or SF3
the derived values of $\log L^*$ typically change by $-0.17$ and
$0.14$, respectively, essentially the same for short and long bursts.
 
Assuming that the BATSE values of $\log L^*$ are offset by $+0.08$,
we find that for our given assumptions about the shape and the
evolution of the luminosity function, long bursts with $T_{90} > 2$ s 
and short bursts with $T_{90} < 2$ s have the same value of the 
characteristic peak luminosity 
$L^* \sim 0.6 \times 10^{51}$ erg (0.064s)$^{-1}$.
As a consequence, short bursts will have lower radiated energies 
than long bursts, which may have a major effect on the afterglows
of short GRBs \citep{pan01}. Among the short bursts there is no 
evidence for any substantial change in $L^*$ for durations as 
short as 0.25 s.

\begin{deluxetable}{ccccccccc}
\footnotesize
\tablecaption{Characteristic Luminosity and Space Density From Various
 GRB Samples\tablenotemark{a}. \label{tbl-1}}
\tablewidth{0pt}
\tablehead{
\colhead{Sample} & \colhead{$\Delta t$} & 
\colhead{$T_{90}$}  & \colhead{n}  &
\colhead{$<V/V_{\rm max}>$} &
\colhead{$\alpha$}  & \colhead{$f_{lim}$\tablenotemark{b}} &
\colhead{$\log L^*$\tablenotemark{c}} & 
\colhead{$\rho_o$\tablenotemark{d}} 
}
\startdata
BD2   & 1024 ms &     &1391 & 0.336$\pm$0.008&$-1.6$& 0.25 & 50.67 & 0.339\\
BATSE & 1024 ms &     & 612 & 0.312$\pm$0.012&$-1.6$& 0.25 & 50.88 & 0.240\\
BATSE & 1024 ms &$>2$s& 469 & 0.291$\pm$0.013&$-1.6$& 0.25 & 51.05 & 0.147\\
BATSE &  64 ms  &$>2$s& 323 & 0.376$\pm$0.016&$-1.6$& 1.00 & 50.72 & 0.250\\
BATSE &  64 ms  &$<2$s& 141 & 0.354$\pm$0.023&$-1.1$& 1.00 & 50.92 & 0.075\\
BATSE &  64 ms  &$<0.5$s& 91& 0.361$\pm$0.029&$-1.1$& 1.00 & 50.86 & 0.040\\
BATSE &  64 ms  &$<0.25$s& 49&0.347$\pm$0.042&$-1.1$& 1.00 & 50.98 & 0.023\\
\enddata

\tablenotetext{a}{We use $H_0 = 65~$ km s$^{-1}$ Mpc$^{-1}$, 
$\Omega_M = 0.3$, and $\Omega_{\Lambda} = 0.7$.}
\tablenotetext{b}{$f_{lim}$ is the limiting flux in 
ph cm$^{-2}$ s$^{-1}$ over the energy range $50-300$ keV.}
\tablenotetext{c}{$L^*$ is the isotropic-equivalent characteristic
peak luminosity in erg (0.064s)$^{-1}$ in the $50-300$ keV band.}
\tablenotetext{d}{$\rho_o$ is the local ($z=0$) GRB rate, in units of
Gpc$^{-3}$ yr$^{-1}$.}

\end{deluxetable}


\begin{thebibliography}{}

\bibitem[Fenimore \& Ramirez-Ruiz(2000)]{fen00}
 Fenimore, E. E., \& Ramirez-Ruiz, E. 2000, \apj, submitted 
 (astro-ph/0004176)
\bibitem[Fishman et al.(1989)]{fis89} Fishman, G. J. et al. 1989,
 in {\it GRO} Science Workshop Proc., ed. W. N. Johnson
 (Greenbelt:NASA), 2-39
\bibitem[Frail et al.(2001)]{fra01} Frail, D. A. et al. 2001,
 submitted (astro-ph/0102282)
\bibitem[Gandolfi et al.(2000)]{gan00} Gandolfi, G. et al. 2000,
 in AIP Conf. Proc. 526, Gamma-Ray Bursts, ed. R. M. Kippen,
 R. S. Mallozzi, \& G. J. Fishman  (New York: AIP), 23
\bibitem[Hakkila et al.(2000)]{hak00} Hakkila, J., Meegan, C. A., 
 Pendleton, G. N., Malozzi, R. S., Haglin, D. J. \& Roiger, R. J.
 2000, in AIP Conf. Proc. 526, Gamma-Ray Bursts, ed. 
 R. M. Kippen, R. S. Mallozzi, \& G. J. Fishman (New York: AIP), 48
\bibitem[Kouveliotou et al.(1993)]{kou93} Kouveliotou, C.,
 Meegan, C. A., Fishman, G. J., Bhat, N. P., Briggs, M. S., Koshut,
 T. M., Paciesas, W. S., \& Pendleton, G. N. 1993, \apj, 413, L101
\bibitem[Mao, Narayan, \& Piran(1994)]{mao94} Mao, S. M., Narayan, R.,
 \& Piran, T. 1994, \apj, 420, 171
\bibitem[Meegan et al.(1998)]{mee98} Meegan, C. A. et al. 1998,
 in AIP Conf. Proc. 428, Gamma-Ray Bursts, ed. C. A. Meegan,
 R. D. Preece, \& T. M. Koshut (New York: AIP), 3
\bibitem[Norris, Marani \& Bonnell(2000)]{nor00} Norris, J. P., Marani, 
 G. F., \& Bonnell, J. T. 2000, \apj, 534, 248
\bibitem[Norris, Scargle \& Bonnell(2001)]{nor01} Norris, J. P.,
 Scargle, J. D., \& Bonnell, J. T. 2001, 2nd Rome workshop,
 (astro-ph/0105108)
\bibitem[Panaitescu, Kumar, \& Narayan(2001)]{pan01} Panaitescu, A.,
 Kumar, P., \& Narayan, R. 2001, \apj, submitted (astro-ph/0108132)
\bibitem[Porciani \& Madau(2001)]{por01} Porciani, C., \& Madau, P.
 2001, \apj, 548, 522 
\bibitem[Schmidt(1968)]{sch68} Schmidt, M., 1968, \apj, 151, 393
\bibitem[Schmidt(1999a)]{sch99a} Schmidt, M., 1999a, \aap Suppl, 138, 409
\bibitem[Schmidt(1999b)]{sch99b} Schmidt, M., 1999b, \apjl, 523, L117
\bibitem[Schmidt(2001)]{sch01} Schmidt, M., 2001, \apj, 552, 36
\bibitem[Steidel et al.(1999)]{ste99} Steidel, C. C., et al. 1999,
 \apj, 519, 1
\bibitem[Tavani(1998)]{tav98} Tavani, M. 1998, \apj, 497, L21

\end{thebibliography}
\end{document}